# Personal Semantic Web Through A Space Based Computing Environment

Ian Oliver, Jukka Honkola

SmartSpaces Laboratory
Nokia Research Center
Itämerenkatu 11-13
00180 Helsinki
Finland
`ian.oliver@nokia.com`

**Abstract.** The Semantic Web through technologies such to support the canonical representation information and presenting it to users in a method by which its meaning can be understood or at least communicated and interpreted by all parties. As the Semantic Web evolves into more of a computing platform rather than an information platform more dynamic structures, interactions and behaviours will evolve leading to systems which localise and personalise this Dynamic Semantic Web.

## 1 Introduction

The Semantic Web in its current form represents information that is either static or monotonically changing. It has its roots in the World Wide Web and thus takes this 'web-wide' persona. The information and ontologies used have standard web-wide semantics grounded in common and generally accepted real-world concepts. Deviation from this - at least in the semantic sense is not permitted.

Much of the usage of information - that will invariably become part of the Web - will be personal and/or local; its semantics, structure and adherence to real-world concepts will be grounded by the local users of that information resulting in many localised Semantic Webs.

Tim Berners Lee's 'Giant Global Graph' is the current reality and overriding direction of the Semantic Web at this point in time. We believe that the Semantic Web and this 'Giant Global Graph' will coalesce into into both larger static entities and smaller, more personal, highly dynamic entities all of which will contribute to a 'Giant Web of Global Graphs'. These individual graphs will vary with content, stability, personality, personalisation, dynamicity, consistency, visibility, privacy and so on.

These local Semantic Webs will contain information that is highly dynamic, non-mononticaly changing, adhering loosely (if at all) to their stated ontologies or even not to any standardised, written, commonly understood ontology and behave and be reasoned about according localised, non-standard and non-intuitive logics.

The dynamicity and monotonicty of information in and forming the Semantic Web will vary depending upon locality and be organised according to person, usage etc as - what we term - spaces.

In this paper we describe our vision and an architectural concept regarding the development of these small, localised information or knowledge spaces[1] by which persons via autonomous 'agents' interact through control-flow-free mechanisms.

## 2   An Architectural Vision

At the time of writing there are over 2 billion[2] mobile devices in use in the World many of which have significant memory and processing power but more importantly provide a mechanism for ubiquitous computing. Augumenting this are the standard fixed installations of PCs, servers as well as consumer devices such as TVs, Media Centers etc. All of these devices could participate in ubiquitous communication.

One of the barriers to interoperability is that different devices apply different formats of information in order to understand information. Technologies such as XML, RDF, OWL etc provide methods allowing us to focus on the semantics of the information rather than the syntactical representation [6,21]. Organising communication between these devices and the applications that run on these devices is known to be hard [2]. A straightforward approach: give a set of devices, a blackboard[3] can be used to *share* information between these devices rather than have the devices explicity send messages to each other. In addition to this if this information is stored conforming to some ontoligical representation then it becomes possible to share information between devices that do not share the same common representation model but rather focus on the semantics of that information.

---

[1] Just 'space' or 'SmartSpace' are our chosen terms in this paper
[2] $2 \times 10^9$
[3] whiteboard,greenboard,billboard,chalkboard...

We propose and have developed a simple system called **Sedvice** that takes the agent, blackboard and publish/subscribe concepts [7,8,20,12,15] and reimplements it in a lightweight manner suitable for small, mobile devices. These agents, which we term 'knowledge processors (KP's)' operate autonomously and anonymously by *sharing* information through spaces. These spaces contain a store of information to which the KPs have access to and can change at will, processing capabilities for reasoning, modifying and analysing that information as well as the usual gamete of security and policy functions.

The Sedvice system was developed with the premiss that information would be shared through localised (potentially) personal spaces; the information would be highly dynamic, ie: changing, and have a semantics primary given through that information's interpretation by any given KP. No control-flow mechanism was to be used but KPs could share information about how external communication, synchronisation and coordination might be achieved.

## 3  Sedvice Architectural Overview

The Sedvice architecture at its simplest is a publish/subscribe system (cf: [7]) consisting of distributed KPs - which consist of user-interface, the logic or internal workings of that knowledge processor and nodes - and then Spaces which consist of information brokers (known as SIBs), an information store and reasoners respectively. Figure 1 shows the relationship between these parts.

A KP is a composite structure which runs on a single device, eg: mobile device, PC, sensor etc. A device may of course run many KPs dependent upon operating system, memory etc. User interfaces are not necessarily necessary or may be extremely simple in nature, eg: LCD display, single button. The Node is the part of the KP which contains all the logic and functionality to connect to some Space: the logic for parsing messages and pointers to subscription handlers between the KP and Space. A node may potentially connect to more than one Space at a time thus distributing and syncrhonising the operations across all connected Spaces or alternatively a KP might contain more than one node.

The basic functionality for manipulating information by a node is:

**Insert:** insert information atomically
**Retract:** remove information atomically
**Query:** synchronously (blocking) query; returns information
**Subscribe:** asynchronously (persistent, non-blocking) set up a subscription for a given query.

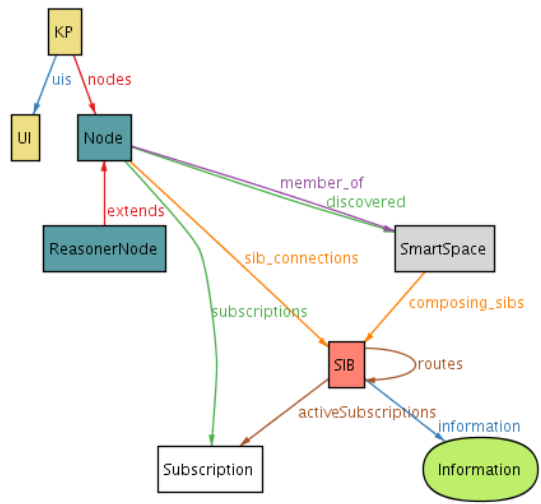

**Fig. 1.** Sedvice SmartSpace Architecture

**Unsubscribe:** terminate processing of the given subscription

A Space at its simplest contains a single method for listening on some transport protocol (eg: TCP socket interface), the logic for processing messages, policy/sercuity and handling subscriptions and finally the information store itself. Additionally a set of 'reasoners' may also be present which are effectively Nodes operating in a restricted environment.

These Reasoner Nodes run after all the pending requests for insertion and retraction of information and process the information in the information store. This processing may be truth maintenance, belief revision, information consistency management, information creation or any other atomic processing required by that space. Reasoner Nodes are scheduled according to priority classes with the reasoners in each class running concurrently and all completing after the next priority class is scheduled. The reasoning for priority classes of reasoners is to control the ordering if two reasoners would interact in unpredictable ways if run concurrently. A reasoner node may be scheduled to run more than once in one whole reasoning cycle if in two different priority classes.

Implementation of the above has been made on various devices - the main requirement being that these devices support the runtime environments we are using: Python, Java (both MIDP and SE), C and OpenC (for Symbian). Current implementations run on Mobile Devices (including: N800/810, N95) and

PCs (Unix, Linux, Windows) - and example test setup is shown in figure 2. Some work is underway to run the KPs on sensors etc. Communication is made over TCP/IP and HTTP protocols which of course can be used over ethernet, GPRS and 3G transports.

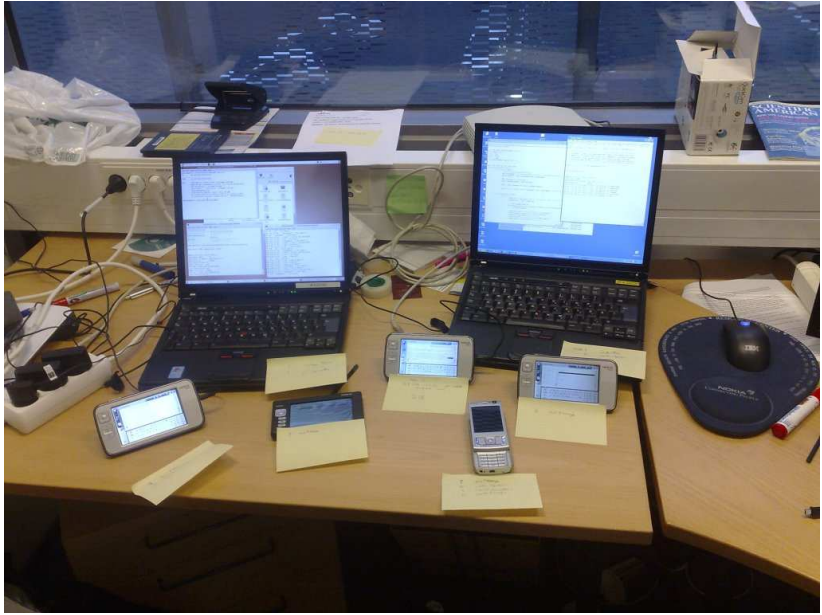

**Fig. 2.** Example Test Setup

### 3.1 Information

The SIB stores the information as a graph and conforms to the rules of the Resource Description Framework (RDF). We use the Wilbur RDF store [10] for storage of this graph.

The basic operation upon the information store are insertion of a graph, retraction of a graph, querying and subscription for information. Insertion and retractions may be combined into a single transactional structure in order to admit atomic update through the atomic application of a retract and insert. The current model theory of RDF only considers the consistency of RDF and not the operations that can be performed upon it; those such as retraction are known to be problematical [5,14] and we take the point of view that both insertion and retraction in our system be based upon the simplest semantics:

$$\Gamma, insert(\phi) \vdash \phi \in \Gamma$$

$$\Gamma, retract(\phi) \vdash \phi \notin \Gamma$$

all other forms of these are refinements of the above. When combining insertions and retractions, while Sedvice allows other semantics to be plugged-in, we take in the simplest case when processing the transaction before applying it to the information store the following re-write rules:

$$\Gamma, insert(\phi), insert(\phi) \vdash \Gamma, insert(\phi)$$

$$\Gamma, retract(\phi), retract(\phi) \vdash \Gamma, retract(\phi)$$

$$\Gamma, insert(\phi), retract(\phi) \vdash \Gamma$$

Of course this will change in the situations involving trust [19] and uncertainty [13].

Queries are synchronous in nature while subscriptions are persistent and asynchronous and whose lifetime is governed by the creating node. In the current system two types of query format are admitted: triple queries and WQL queries. Triple queries are effectively pattern matches over the RDF triple while WQL is a lisp-like path based query language. One important difference is that Wilbur's static reasoning engine [11] only runs with WQL queries - this engine is described in and computes a deductive closure over certain RDF constructs such as the subtype hierarchies and `owl:sameAs` [4].

No attempt is made by the Space to enforce consistency or integrtity of information according to the stated ontologies (through the RDF typeOf relation); internal reasoning nodes may be present which performs this activity if the space has been instantiated that way. Information is explicitly semi-structured and may take on any form that the KPs insert/retract.

While typing constructs and namespaces are present (if this information is inserted of course) these do not necessarily mean that a node querying for that information will necessarily interpret that information according to the implied ontology. The semantics of the information is interpreted by the reader, merely implied by the writer and grounded in the real-world context of the node - any two given KP's may disagree about this.

---

[4] Lassila: Identity Crisis and Serendipity

### 3.2 Other Functionality

The Space provides further functionalities regarding the joining and leaving KP/Nodes and policy management. KPs have a set of credentials which are passed during the join operation

The dual of the KP/node instantiated leave and join operations are the Space instantiated invite and remove operations. These operations are not necessarily provided by every Space nor understood by every KP/Node.

Connectivity is provided through a set of 'listeners' which provide access via any given specified protocol. Typically used is TCP/IP through the standard sockets mechanism but a Bluetooth based listener or one that uses HTTP/S have also been developed. Listeners can provide preprocessing of the incomming messages if necessary; for example with Bluetooth profiles. Any number of listeners may be provided at any time (at least one is necessary however!)

A Space is actually constructed from a number (at least one) 'Semantic Information Brokers' (SIB) which contain the aforementioned schedulers, management information, listeners and information store. A Space is represented by these SIBS: the total possible connectivity methods is given by the distributed union of all the representing SIB listeners.

SIBs communicate internally to ensure that membership and credentials of the KP/Nodes that have joined that Space - they may connect via any listener and any SIB in that Space. From the point of view of any KP, the information available is always the distributed union over the transititve closure of the routes between all the SIBs - the SIBs contain routing tables to other SIBs and within a Space all the SIBs are *totally routable* but not necessarily totally connected.

## 4 Applications

The traditional notion of the monolithic, control-oriented, single device application does not exist. We explore two notions of application: the first is how applications are constructed out of KPs and the second is the notion of emergent application.

### 4.1 Application Construction From Knowledge Processors

What the user perceives as being an application is constructed from a number of independently operating KPs which - if they have user-interfaces - may be grouped together visually (or by some other means, eg: audio) so that their combined results may be perceived as a single whole; for example, compare with the notion of Widget in the Nokia Widsets or dashboard application under Apple's OSX and potentially even as Spimes or Blobjects[5]!

We present a simple example of chat functionality where persons (or agents) can send and receive messages. Functionality is divided up in most cases across a single ontology (figure 3) by individual operations, eg: send message, view messages etc. This functionality is embedded inside a KP as visualised in figure 4.

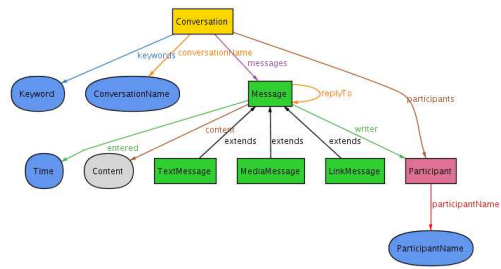

**Fig. 3.** Simple Chat Ontology

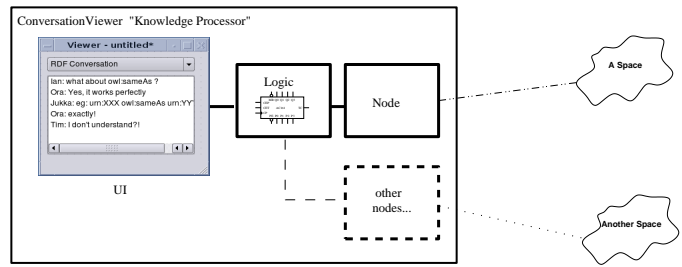

**Fig. 4.** Example Knowledge Processor

---

[5] http://www.viridiandesign.org/notes/401-450/00422_the_spime.html

At any point in time there are a number of instances of these KPs distributed amongst (theoretically) any number of devices. Typically for one (human) user in our chat example we require four KPs: one message viewer, one message writer/sender, one conversation list viewer and one to join and leave conversations. The actual distribution of functionality may vary depending upon implementation.

Specifically the viewer KPs would subscribe to particular parts of the information in the space, for example in WQL[6] one might write the following query to obtain the list of current conversations:[7]

```
ns#Conversation | (:inv !rdf:type)
```

which returns a set or bag of URI's which have are 'of type' ns#Conversation. To obtain the messages in a given conversation (say, one with the URI xyz):

```
xyz | (:seq messages)
```

and then the following to obtain the writer's name, message contents and the uri of the messages to which this is a reply-to (if they exist) in using some pseudocode.

```
foreach m in xyz | (:seq messages):
    writer  := m | (:seq writer)
    content := m | (:seq content)
    replyto := m | (:seq replyTo)
```

This function would be then called everytime the set of results returned by the original subscription changed, ie: when some KP enters a new message into the given conversation.

A meaningful situation for this occurs when there are more than one active participant (obvious!) which entails that each participant has enough KPs in order to fulfil the scenario of having a conversation. This does not mean that each participant - in this cas - would have four KPs, one each of the four types mentioned. It is possible (and desirable) that the user might have a number of each KP distributed amongst their devices to allow for example, viewing of messages to be made in many places. The possibility of having zero KPs of some type is also permissable though this restricts the functionality percevied by that user of the system though situations such as someone wishing only to view messages or which conversations are active is possible.

---

[6] Wilbur Query Language

[7] For simplicity of reading we use ns as a shorthand for some namespace declaration

One caveat of course is the lack of control-flow and synchronisation mechanisms in the space - at least in the chat example this is not a problem as the users of the KPs are not sending messages to each other but rather sharing the contents which may be being read by others. Syncrhonisation and coordination mechanisms need to be build into the ontologies and this is still very much a research question. Of course, KPs may share information about their existance and how to communicate with them outside of the space environment but this is a different issue.

### 4.2 Emergent Functionality

Additional functionality can be created by adding KPs which interpret and modify the information that is being shared amongst the currently running KPs. Typically this is made by KPs which link two or more information structures together. Consider the situation in the RDF graph in figure 5 where we have both a message (in some conversation) and a weather report.

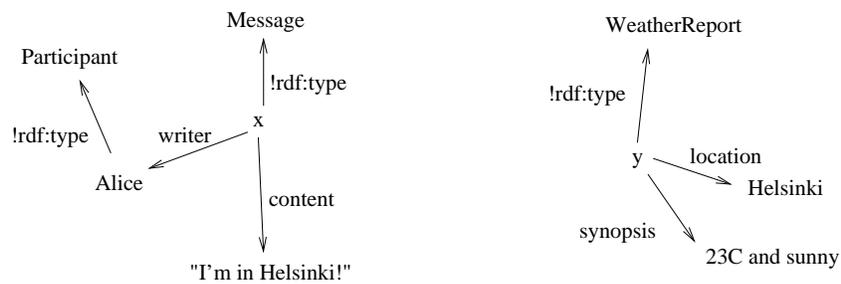

**Fig. 5.** Example Message and Weather Report as an RDF Graph

A suitable subscribe message viewer KP would react to this graph and probably return `Alice: ''I'm in Helsinki''` on the user-interface.

A KP could be constructed such that it subscribes also to new messages and examines their contents for place names gathered from a subscription to locations in weather reports. Upon detecting a match rather than generating a new set of nodes in the graph it could simply link these together such as the uri of the weather report is also a message in a conversation - as shown in figure convgraph2.

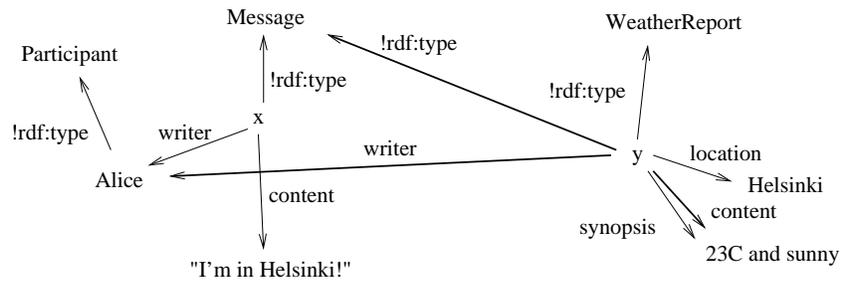

**Fig. 6.** Example Message and Weather Report as an RDF Graph (2)

Then simply the KP that has subscribed for messages then outputs the text `23C and sunny`. Of course there are timing issues and the writer of the KP might take these into consideration through the use of the reply-to mechanism in the conversation ontology or through more advanced construction of the contents of the message.

This kind of linking information together is the key to the semi-structured form of the RDF based Semantic Web and thus key to the methods in which the additional functionality of the chat application in this example is perceived by the users. No changes are required of the KPs relating to the chat nor of any of those dealing with gathering and generating the weather information. There are problems in that KPs could be destrictive in their behaviour and at present we are investigating mechanisms to prevent this at both the information and KP levels.

## 5 Inconsistency, Incompleteness and Semantics

Information is stored using RDF as graphs made of RDF triples. Some items of information *can* be expressed as a single triple but normally as a set of triples which forms one or more RDF graphs - a single triple is the smallest possible RDF graph. We talk about information meaning any collection of RDF graphs [8].

The semantics of the information is implied by the writer by way of what typing information is provided, the amount of information and what ontologies are used. The reader of any piece of information interprets the information according to its own criteria which is influenced by the typing, ontologies etc [3]. However there is nothing stopping the reader from completely misinterpreting the information; we do not see this as a problem but rather a way by which dynamicity and semantics of the whole information set can evolve [22,1]. More accepted ontologies and information will imply more rigid interpretation.

---

[8] An analogy we use is that triples are letters, RDF graphs are words - there are not many single letter words, thus generally we must talk about graphs and not triples

Our system does not enforce any notion of consistency and completeness according to any given ontology. When information is inserted or retracted no checking against any ontology is made. If checking is required then we summise that this is provided by a KP/Node or better, a Reasoner Node, which enforces this through belief revision.

Without consistency checking this now allows us to admit models that are incomplete in the sense that some information is not known or just not presented by the writer. The interpretation of whether the information is not known or not specified is decided locally by some reader.

Further from this we have already noted the dyanmicity of information and the potential for inconsistency of that information with respect to the ontologies to which the information was structured according to. Over time the ontologies themselves will change, albeit at a slower rate than the information itself; this will happen in one of two ways, either realignment into either more strict ontologies or more 'correct' with respect to the changes in the information structure they are representing, *or*, the ontologies will become degenerate structures whicih admit many-to-many relationships between their concepts. The reality is that over time both realignment and degeneration will occur cyclically.

However, basing everying upon ontologies misses the point that the concepts and their structures defined in the ontology actually refer to real-world concepts - this is a deeper level of semantics which so far has not been touched upon in current systems. This requires more work especially in the context and with relationship to systems that are built upon potentially highly varying semantic interpretation.

# 6 Conclusions and Future Work

This paper elaborates upon our current thinking, vision and implementation of a system for information sharing using the Semantic Web along with a simple example of its usage. We believe the 'web' will become more of an information push and share environment rather than the current 'pull' of information. Infor-mation will coalesce into locally shared spaces which themselves form a web of spaces.

Work is now progressing into exploring how reasoning over rapidly changing information sets and we are focussing particularly on non-monotonicity and uncertanty [4,9]. Coupled with this is the development of ontological structures and the generation and distribution of KPs from those ontological structures. Furthermore as we have no explicit policy and trust mechansisms work also continue in this field [23].

Eventually the Semantic Web will emerge as *the ubiquitous computing platform* [18,17] in its own right and technologies such as those described here either in implementation or basis of ideas and vision are steps towards this new paradigm of computing *in* the Web.